**Mohammadreza Ahmadi**

# Tissue Engineering and Regeneration of Skin and Hair Follicle Growth From Stem Cells

## INTRODUCTION

Many people with skin diseases such as chronic wounds, non-healing and diabetic ulcers need reconstruction and regeneration of their skin. In addition, the medical industry also needed a method of skin rejuvenation and reconstruction for cosmetic purposes, even for healthy people. Reconstructive medicine used the method to deliver pluripotent stem cells to the targeted tissue.

33 years after the introduction of bone marrow stem cells, fat-derived stem cells have become an excellent source for cell therapy. In 1961, two Canadian scientists first introduced stem cells. These cells, later found to be hematopoietic stem cells, have been used successfully to treat leukemia and some severe autoimmune diseases called bone marrow transplants. In 1968, another stem cell was introduced into the bone marrow, which has been shown to be effective due to its high ability to regulate immunity in many diseases, including skin, bone, joint diseases, heart, brain, nerves, and kidney. Nevertheless, it took scientists another 33 years to discover that human adipose tissue also has a stem cell, similar to bone marrow stem cells. Each gram of human fat has 100 to 500 times more stem cells than bone marrow and is easier to obtain. Although there are differences in bone marrow stem cells and fat, these advantages have attracted the attention of many researchers.

Mesenchymal stem cells are the most interesting because when they get placed in the right environment and are coupled with the proper growth factors, they can choose a pathway to differentiate into that targeted tissue. Also, they are very accessible, inexpensive, easy to extract, and reproducible. The sources of these mesenchymal stem cells are bone marrow, bone, connective tissues, fats, etc. The best reservoir that has recently been discovered for mesenchymal cells is fat. Fats are made out of fibrous, collagen fibers, and fibroblasts.

On one hand, by injecting collagenase enzymes into the fat, all the collagen fibers and fibroblasts would slip away, and the pure mesenchymal stem cells would be available. On the other hand, In the engineered stem cell vascular fraction (SVF), which is a pellet containing mesenchymal cells, growth factors, inflammatory cells like macrophage-monocyte and fibroblasts, the mesenchymal cells get separated from all the other components by the centrifugation process. Placing the mesenchymal stem cells into the targeted environment and providing the proper growth factors can produce fibroblasts and collagen to reconstruct the skin.

As mentioned above, the stem cell environment is critical, and there is a possibility of stem cell movement due to the blood flow and paracrine, which the stem cell would secrete and affect its environment. Therefore, the matrix that keeps the stem cell in its place from getting flushed away by the blood flow and affecting its environment, known as the scaffold, was discovered. The scaffold has to have specific characteristics that would avoid a foreign body reaction; It could not be antigenic ,allergic, or infectious. Additionally, it needed to be accessible and cheap to produce. The newest method utilized has been using alginate scaffolds, containing the mesenchymal cells and the essential growth factors, into the targeted cell resulting in stimulating the CD44 receptors on the skin's fibroblasts, causing a rush of collagen production. As we learned in our lectures, the green fluorescent protein bone marrow was implanted into non-green fluorescent protein mice resulting from HCT, (hematopoietic cell transplantation) improved the animal survival in a mouse model, which was surprising.

**Focused Tissue**

In this paper, the tissue regeneration of skin and the probability of hair growth from stem cells are being investigated. Over time, The biological skin replacement has dramatically improved from individual application of skin cells to cells and scaffold combinations for treatments, healing, and acute and chronic skin wounds. Permanent skin substitutes with cellular engineering combination has shown to have crucial wound closure of more than 90% of the total body surface region in burns [7]. As an advantage the morbidity and mortality of acute and chronic wounds have brought lower; However, yet they have been unable to eliminate all skin mechanisms and functions [7]. Therefore, Hypopigmentation, the absence of lymphatic networks and stable vascular, hair follicles, sweat glands, adipose, and inadequate innervation are some of the missing elements of cellular or biological skin replacements equation [7]. The complete restoration of the anatomy and physiology of uninjured skin is highly dependent on the modulation of biological pathways of embryonic and fetal formation. Understanding and incorporating developmental biology into possible biological skin replacements can have a full recovery of anatomy and physiology, along with eliminating morbidity from skin wounds and scars.

In recent years, improvements in early wound excision, respiratory assistance, fluid revival, treatment of inhalation damage, enhanced immune function, infection prevention, integration of aerobic activity during healing, and advancement of anti-scarring techniques have all been accomplished in burn care. These advancements also resulted in substantial decreases in death, hospitalization, and long-term morbidity. Many strategies for healing deep partial and full-thickness burns have been developed by researchers, including the placement of the epidermal barrier through autologous or allogeneic transplantation. Autologous cells can permanently reside on the wound to have definitive wound closure. In contrast, allogeneic cells can stay on the wound for a few days or weeks; However, the combination of the allogeneic fibroblast with biodegradable scaffolds would provide extracellular matrix and growth factors components would facilitates faster wound closure by motivating autologous healing, but it does not last long maybe for only a few days to weeks.

**Potential Issues**

Although these methods do tend to decrease mortality in major burns, after transplantation, they neglect hair follicles and glands. Scientists use techniques which can damage cells such as multi-layering of a variety of cells and the transition of cell-matrix droplets onto a substrate known as "ink-jet printing," by amplifying a laser pulse. The brief exposures to high pressure, temperature, or biochemical toxins would damage the cells [7].

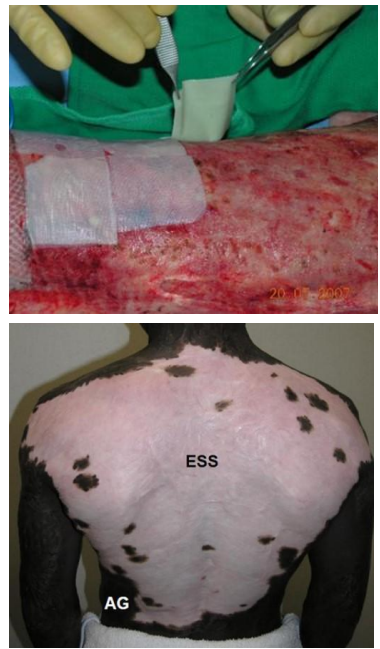

**Figure 1**. The clinical application of autologous engineered skin substitutes.

**Potentially Overcome Solution**

Regenerative medicine is meant to restore the functionality of a harmed,

malfunctioning, or absent tissue. In regenerative medicine, there are three primary methods. Cell-based therapies are the first approach executed by the cells that are delivered to either directly regenerate a tissue or through the paracrine functions. The second method, the material-based methods focuses on biodegradable materials that are often fabricated with cellular functions. The most important method, that is the main focus of this paper, is tissue engineering: combining the use of cells and a biodegradable scaffold to construct a tissue [4].

At the base of hair follicle cells ADSCs (adipose tissue-related stem cells) are mesenchymal stem cells found in subcutaneous tissue. Controlling skin regeneration, aging, and structural deficits are significant roles of these cells [8].

Proliferating and dividing into skin cells to replace damaged or dead cells are known to be done by ADSCs. Additionally, they also stimulate cell replication and the healing process through an autocrine and paracrine pathway by stimulating CD44 fibroblast receptors. During wound recovery, ADSCs have a high capacity to migrate to wounded sites and are recruited quickly, in addition to their division into dermal fibroblasts (DF), endothelial cells, and keratinocytes. Adipose-derived stem cells are primarily more proliferative and have immunoregulatory properties, which means they are isolated in a less disruptive and more reproducible manner with the capability of inactivating T cells [8]. Compared to bone marrow, ADSCs have higher multipotency, which is actually needed more for ectodermal and endodermal tissue repair. ADSCs will secrete a diverse secretome, resulting in increased cell proliferation and differentiation, migration, and enhanced cellular and microenvironment safety. A set of trophic factors that secretome correlates to cytokines, growth factors, and chemokines that enable ADSCs to function as paracrine tools rather than cell replacement.

ADSCs were also discovered in subcutaneous tissue. This involvement plays an essential role in skin restoration and regeneration by retaining and rejuvenating skin tissue structure (even it can be as a physiological reaction to local injury) by seeding younger cells to the epidermis, which they later differentiate into dermal fibroblasts, keratinocytes, and other skin components [8].

The tendency of different epithelial cells to secrete growth factors and extracellular matrix (ECM) proteins maintains a differentiation of rejuvenation and regeneration in the ADSC microenvironment. Furthermore, the primary sources of extracellular matrix (ECM) are ADSCs and DFs proteins involved in skin structure and functionality. Extracellular vesicles are lipid bilayer vesicles that include several RNA organisms (including mRNAs and miRNAs), soluble (cytosolic) proteins, and transmembrane proteins in the proper and functional orientation [8]. EVs are involved in various mechanisms, including intercellular connectivity, membrane protein, lipid recycling, immune regulation, apoptosis, angiogenesis, cellular proliferation, differentiation, and tissue engineering to modulate cell recruitment.

Exosomes, which are emitted through the fusion of the multivesicular structure with both microvesicles and the cell membrane that have been released directly from the cell membrane, are examples of extracellular vesicles (EV) [4]. The many processes in which EVs are active and the elasticity in which they may affect the activity of recipient cells allows EV, a fascinating subject, to be utilized for both therapeutic and diagnostic purposes. EVs are implicated in tissue and organ injury reconstruction and can clarify any paracrine symptoms seen in stem cell-based therapeutic approaches. MSCs (multipotent mesenchymal stem cells) are discovered as residual stem cells in majority adult organs, especially adipose tissue. These cells exhibit standard mesenchymal cell properties in vitro and are segregated from the mesenchymal vascular fraction.

Extracellular vesicles have a paracrine impact on cell morphology, migration, proliferation, and differentiation. These paracrine results of EV can be helpful in regenerative medicine. EV may be used in regenerative therapy in various ways, like injection, combining with hydrogels, and covering scaffolds with EV utilizing fibrous tissue hydrogel. Intercellular communications are essential roles of EVs through secretome that engage in. Thus, their content and functionality can vary based on the conditions of the vesicular generating cells. Changes in EV content in response to conditions specific to growth would play a crucial role in these cases by affecting tissue regeneration and wound repair by identifying new pathways into intercellular signaling.

Epidermolysis bullosa is a genodermatosis, a hereditary skin disease that usually mutations of genes encoding extracellular matrix proteins (collagen) are the primary cause. Severe and prolonged skin blistering with extreme and potentially fatal complications are examples of this disorder. There is no treatment for epidermolysis bullosa at the moment. However, enabling autologous therapies developed from a patient-specific renewable community of cells count as advantages in gene therapy and editing methods, along with the development of pluripotent stem cells from patients with epidermolysis bullosa [7].

Another example of skin disease is diabetic foot ulcers, known as (DFU), which is a significant difficulty that often results in infection or even death. The mesenchymal stem cell-based therapeutic method has shown effectiveness in wound treatment, presenting a theoretically systematic alternative by treating various microaerobic factors.

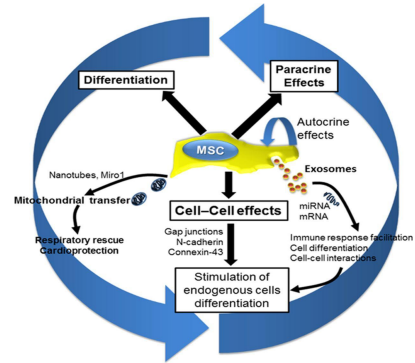

**Figure 2**. The overall combinational therapy.

**Base Material and The Interaction Mechanism**

Damaged tissue repair entails the existence of cells capable of repairing the damaged tissue and the presence of a microenvironment that facilitates adequate tissue regeneration. Furthermore, cells must be directed to form a structure of the suitable size, shape, and in certain situations, structural support. In damaged tissue, the extracellular metric is often damaged, functionally disabled, or missing, while In healthy tissue, the ECM guides and regulates these processes.

Scaffolds are structures that include the conditions for cell retention and tissue regeneration and are used to solve this problem. Scaffolds may be either natural (decellularized ECM or enhanced elastin - or collagen gels) or synthetic (synthetic hydrogels or porous polymer scaffolds) [4].

To enable suitable cell populations, artificial porous scaffolds must meet precise criteria. A synthetic scaffold, ideally, offers temporary protection microenvironment, biodegradable, and is gradually substituted by autologous ECM. By providing a porous structure required, cells would be able to move or they can be seeded in the scaffold to provide an environment with a sufficient supply of nutrients [4]. Cellular components are removed from the tissue using a mixture of enzymatic, physical, and chemical therapies to reduce the possibility of immune responses against antigens and the spread of pathogens in these scaffolds.

Electrospinning is the most widely utilized technique for creating porous synthetic scaffolds since it allows for the development of structures of complicated geometry variations of fiber forms in both mixed and layered configurations. Polycaprolactone (PCL), polyglycolic acid (PGA), polyhydroxyalkanoate (PHA), and polylactic acid (PLA) are biodegradable polymers used in electrospinning.

Combining fiber types in complex patterns, scaffold degradability, durability, and biological activity can be modulated. MSC migration, proliferation, and osteogenic differentiation are improved by coating the scaffold with collagen peptide as an example. Scaffolds may also be engineered to release proteins, peptides, or cytokines through degradation or coating fibers with a biodegradable material like fibrin or gelatin containing a combination of these bioactive molecules. Also, because of their heavily hydrated structure, low toxicity, and proximity to the extracellular matrix, hydrogels are the most commonly used bio-inks in tissue engineering to build scaffolds. As a result, bio-ink built with gelatin and natural polymers is compatible with various soft tissue cell types and can be combined with biological molecules such as proteins, growth factors, and pharmaceutical molecules.

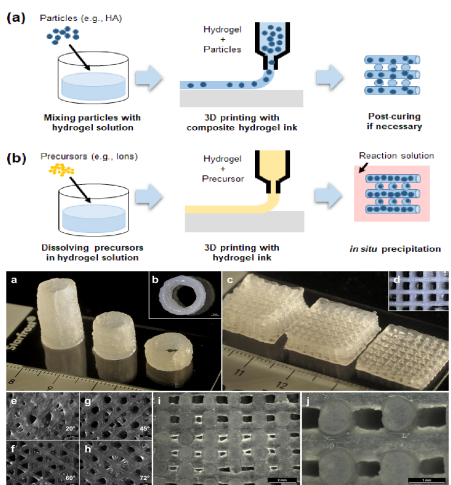

**Figure 4**. The Bioink hydrogel scaffold madel.

## CONCLUSION

Many people suffering from skin disorders such as chronic wounds, non-healing ulcers, and diabetic ulcers need skin repair and regeneration. Aside from the diseases listed above, the industry needed a skin rejuvenation system and regeneration for cosmetic purposes. The procedure used to deliver pluripotent stem cells to the desired tissue was known as reconstructive medicine. Mesenchymal stem cells are the most fascinating since, when put in the correct setting and stimulated with the proper growth factors, they could choose a pathway to differentiate into the desired tissue. They are also very available, inexpensive, simple to extract, and reproducible. These mesenchymal stem cells are derived from bone marrow, bone, connective tissues, fats, and other tissues. Fat is the ideal reservoir for mesenchymal cells that have recently been identified. Fibrous, collagen fibers and fibroblasts create fat.

On one side, by injecting collagenase enzymes into the fat, all of the collagen fibers and fibroblasts can be removed, leaving just the pure mesenchymal stem cells. In the engineered stem cell vascular fraction (SVF), a pellet comprising mesenchymal cells, growth factors, inflammatory cells such as macrophage-monocyte, and fibroblasts, the mesenchymal cells are segregated from the other components through centrifugation. Placing mesenchymal stem cells in the target environment and supplying the required growth factors can result in the formation of fibroblasts, the production of collagen, and the reconstruction of the skin. As previously stated, the stem cell condition is essential, as there is a risk of movement of the stem cell due to blood circulation and paracrine, which the stem cell will secrete and influence its environment. As a result, the scaffold is the structure that prevents the stem cell from being flushed away by the blood supply and disrupting the environment. The scaffold must have unique properties that prevent foreign body reactions, are not antigenic or allergic, and do not induce infection. It must also be affordable and readily available. Alginate scaffolds have been used in the most recent process. The scaffold containing the mesenchymal and critical growth factors into the desired cell will, for example, activate the CD44 receptors on the skin's fibroblasts, increasing collagen output.

      In this matter I genuinely believe that this approach would expedite and help the reproduction and rejuvenation of skin cells by the help of the mesenchymal stems cells, and alginate scaffolds.